\documentclass[letterpaper, 10 pt, conference]{IEEEtran}  

\usepackage{graphicx}
\usepackage{hyperref}  
\hypersetup{bookmarksopen=true} 
\usepackage{float} 
\usepackage{amsmath}
\usepackage{amssymb}
\usepackage{color}
\usepackage{multirow}
\usepackage{atbegshi}
\AtBeginDocument{\AtBeginShipoutNext{\AtBeginShipoutDiscard}}

\IEEEoverridecommandlockouts 

\IEEEoverridecommandlockouts
\IEEEpubid{\makebox[\columnwidth]{~
\copyright~2018 IEEE \hfill} \hspace{\columnsep}\makebox[\columnwidth]{}} 

\begin{document}

\title{\LARGE \bf
Parallelized Linear Classification \\
with Volumetric Chemical Perceptrons
}

\author{
    \IEEEauthorblockN{
        Christopher E. Arcadia$^\dagger$, 
        Hokchhay Tann$^\dagger$, 
        Amanda Dombroski$^\ddagger$, 
        Kady Ferguson$^\ddagger$,
        Shui Ling Chen$^\ddagger$,\\
        Eunsuk Kim$^\ddagger$, 
        Christopher Rose$^\dagger$, 
        Brenda M. Rubenstein$^\ddagger$, 
        Sherief Reda$^\dagger$, 
        and Jacob K. Rosenstein$^\dagger$
    }
    \IEEEauthorblockA{$^\dagger$School of Engineering  ~~~~~~   $^\ddagger$Department of Chemistry}
    \IEEEauthorblockA{Brown University, Providence, RI 02912, USA}
}

\thanks{Correspondence: jacob\_rosenstein@brown.edu \\}

\thanks{This research was supported by funding from the Defense Advanced Research Projects Agency (DARPA). The views, opinions and/or findings expressed are those of the authors and should not be interpreted as representing the official views or policies of the Department of Defense or the U.S. Government.}

\maketitle

\pagestyle{plain} 
\setcounter{page}{1}

\begin{abstract}
In this work, we introduce a new type of linear classifier that is implemented in a chemical form. We propose a novel encoding technique which simultaneously represents multiple datasets in an array of microliter-scale chemical mixtures. Parallel computations on these datasets are performed as robotic liquid handling sequences, whose outputs are analyzed by high-performance liquid chromatography. As a proof of concept, we chemically encode several MNIST images of handwritten digits and demonstrate successful chemical-domain classification of the digits using volumetric perceptrons. We additionally quantify the performance of our method with a larger dataset of binary vectors and compare the experimental measurements against predicted results. Paired with appropriate chemical analysis tools, our approach can work on increasingly parallel datasets. We anticipate that related approaches will be scalable to multilayer neural networks and other more complex algorithms. Much like recent demonstrations of archival data storage in DNA, this work blurs the line between chemical and electrical information systems, and offers early insight into the computational efficiency and massive parallelism which may come with computing in chemical domains.
\end{abstract}



\section{Introduction}
\label{sec:intro}

It is widely appreciated that living systems make use of both electrical and chemical domains for information processing. Cells simultaneously balance numerous internal states and energy gradients, using chemical energy to efficiently maintain both electrical and chemical potentials and interact with their environment \cite{alberts2017}. Neurons, whose complex electrochemical interconnections have inspired modern artificial neural networks \cite{deepNNoverview}, use chemicals for both short-term and long-term memory \cite{goelet1986}. While some of these behaviors have been mimicked in electronic systems, it is impossible for an electrical system to fully recreate the massive parallelism and emergent properties that come from the diversity of subtle molecular interactions and coexistence of thousands of unique chemical compounds. Inspired by these possibilities \cite{RoseISIT,Kennedy2018,molecularneuron,molecularsensing}, we aim to develop a computational framework to concurrently process digital information represented in solutions of chemical compounds. Noting in particular the growing interest in DNA-based data storage \cite{dna-fountain, dna-crispr-cas-encoding, dna-rand-access}, we anticipate that methods of processing chemically stored information will become increasingly relevant.

In this paper, we explore a novel approach to computing with chemical solutions and offer the following contributions towards this goal:

\begin{enumerate}
\item We devise a method to encode binary data into the chemical composition of liquid samples. We additionally show that multiple datasets can be stored in parallel with multiple coexisting chemicals.
\item We use programmable robotic liquid handling sequences to perform volumetric multiply-accumulate operations on parallelized chemical datasets.
\item We utilize high-performance liquid chromatography to read and verify the results of the chemical calculations.
\item We chemically encode several images of handwritten digits from the MNIST database, and implement several single-layer volumetric chemical perceptrons which successfully classify the images. We additionally quantify the performance of our method with a larger set of binary vectors.
\end{enumerate}

The organization of this paper is as follows: We provide the necessary background for this work in  Section \ref{sec:background}. In Section \ref{sec:methods}, we discuss our methods for chemical encoding, computation, and readout. A description of the system we developed to perform these functions is given in Section \ref{sec:development}. In Section \ref{sec:experiments}, we demonstrate linear classification of several MNIST images and Boolean test vectors. In Section \ref{sec:future}, we discuss ways to scale and extend our chemical computation scheme. Section \ref{sec:conclusion} summarizes the main conclusions of this work. This effort is intended as a first step toward realizing general-purpose chemical-domain computational tools.


\section{Background} \label{sec:background}

\subsection{Perceptron Classifier}

A perceptron is a simple linear classifier which can be trained to determine whether or not an input belongs to a certain class \cite{perceptron, perceptron1,prml}. A perceptron uses a set of constant coefficients to compute a weighted sum of input features and thresholds the result to produce a Boolean label. The computation can be written as:

\begin{equation}
z = \sum_{i=1}^N w_i \cdot x_i + b 
\label{eq:weightedsum}
\end{equation}

\noindent where $N$ is the number of input features (e.g. pixels in an image or dimensions of a vector), $x_i$ and $w_i$ are the $i^{th}$ feature and its corresponding weight, and $b$ is a scalar bias. When the bias is nonzero, it is common to fold $b$ into $w$ by introducing an additional input feature whose value is always $1$. The summation result $z$ determines the class label, $\ell$, of the input according to the following threshold:

\begin{equation}
\ell = 
 \begin{cases}
      \text{match,} &\text{if } z > 0\\
      \text{mismatch,} &\text{otherwise}\\
   \end{cases}
   	\label{eq:classdecision}
\end{equation}

For a linearly separable dataset, the training for a perceptron is guaranteed to converge such that a set of weight values will be found that correctly classify all training points. While a perceptron is a simple classifier, it can be used as a building block for larger discriminative models, such as multilayer perceptrons \cite{multilayerperceptron} and support vector machines \cite{SVMnonlinear}, that can realize more complicated decision boundaries. Additionally, the multiply-accumulate (MAC) operations performed in a perceptron are necessary for many pattern classification algorithms. By demonstrating that our proposed method can reliably perform these operations, we intend to show that chemical mixtures can provide an interesting basis for alternative computing. 

\subsection{Mass Conservation in Chemical Mixtures}

It is helpful to consider some physical properties of the simple chemical mixtures that we will use for the proposed computations. If we assume that we are working with well mixed solutions and that there are no chemical reactions in these solutions, then the simplest constraint is the conservation of mass. When two or more volumes containing the same chemical are combined, the final mixture will have a concentration that is a linear function of the concentrations of that chemical in each of the sources. 

More formally, if a mixture of $N$ sources, each containing a concentration $C_i$ of a certain chemical, is formed by transferring a volume $V_i$ from each source to a common destination, then the final concentration will be given as:

\begin{equation}
	C_f = \sum_{i=1}^N  \frac{V_i}{V_f} \cdot C_i 
	\label{eq:mixconcentration}
\end{equation}

\noindent where $V_i \cdot C_i $ is the total mass of the chemical added to the destination and $V_f$ is the final total volume in the destination. The parallels between Equations \ref{eq:weightedsum} \& \ref{eq:mixconcentration} should be apparent. We additionally note that this equation can be extended through linear superposition to mixtures of different chemical species, as long as they do not interact with each other. We take advantage of this simple relationship to form the basis for parallel chemical computation.

\section{Proposed Method for Chemical Computing} \label{sec:methods}

\begin{figure}[t]
\centerline{\includegraphics[scale=0.35]{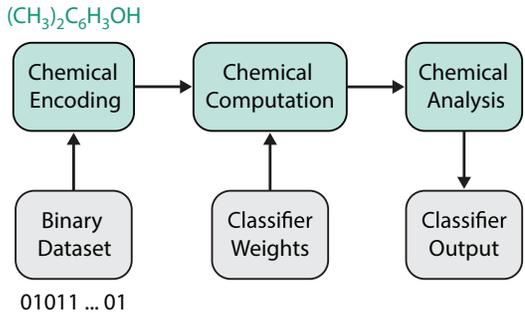}}
\caption{A conceptual block diagram of the chemical computation scheme. Binary datasets are encoded into discretized mixtures of chemicals. Computations can be performed on these chemical mixtures through quantitative sampling, based on the desired classifier's weights, and mixing of their contents. The computation output is initially still in the chemical domain, and can be assessed using analytical chemistry techniques.
}
\label{fig:concept_block_diagram}
\end{figure}

A high-level summary of the proposed computation scheme is shown in Figure \ref{fig:concept_block_diagram}. We begin by encoding the ones and zeros of a binary dataset into a pattern of chemicals in an array of isolated fluid volumes. After translating the data to chemical form, we query the chemical dataset by performing the volumetric MAC operations needed to implement a single-layer perceptron. The chemical output of the MAC stage is analyzed to measure the concentrations of its information-carrying compounds. Finally, we appropriately threshold the concentrations of each compound in the output pools to produce the perceptron's Boolean labels.

In Table \ref{tab:domains}, we relate the parameters of a chemical mixture-based system to familiar electrical terms. A unique advantage of data storage and processing with chemicals is the parallelism that can be achieved by operating with multiple coexisting chemical species. This potential is realized in many biological contexts, such as in bacterial communication \cite{taga2003} and neural signaling \cite{bloom1984}. Another benefit of chemical data storage is its potentially high information density, as has been noted for DNA \cite{dna-next-gen-storage,dna-archival-storage}.

\begin{table}[b]
\renewcommand{\arraystretch}{2}
\centering
\begin{center}
\caption{Analogies between Electrical and Chemical Domains}
\begin{tabular}{|r|c|c|}
\hline 
\textbf{Domain} & \textbf{Electrical} & \textbf{Chemical} \\ \hline
\textbf{Information Carrier} & Electrons & Molecules \\ \hline
\textbf{Digital `1'} & High Voltage & High Concentration \\ \hline
\textbf{Digital `0'} & Low Voltage & Low Concentration \\ \hline
\textbf{Transport Medium} & Conductor & Solvent \\ \hline
\end{tabular} \label{tab:domains}
\end{center}
\end{table}

\subsection{Encoding Data in Chemical Mixtures}

In order to carry out computation in the chemical mixture domain, we need to first create a representation for our data. Since chemically expressed data will be stored in microwell plates, we map each well position to one bit in the input data. The value of the data at a given position is represented by a high concentration (`1') or low concentration (`0') of a designated chemical. This strategy could also be extended to multi-bit concentration coding schemes, with the trade-off of reduced noise margin.

\begin{figure}[t]
\centerline{\includegraphics[scale=0.2]{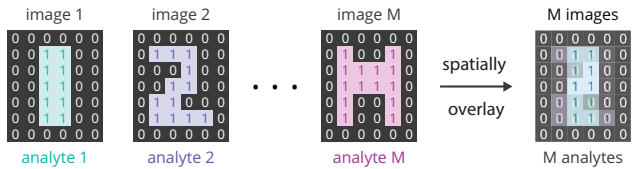}}
\caption{Data is stored in isolated wells containing quantitative chemical mixtures. The concentrations of these chemicals reflect the values of the binary input data. Each bit address in the input data is assigned to one grid location on a microplate, while the value of each bit is encoded in the concentration of a particular chemical compound at that position. Multiple datasets can be simultaneously stored in the same fluid containers by using multiple distinct chemicals.}
\label{fig:concept_storage}
\end{figure}

To enable parallel data storage and processing, we can take advantage of the diversity of chemical compounds and overlay (concurrently encode) features from multiple input datasets in the same set of microplate wells. For instance, we could take many binary images and realize all pixels with the same position in a single well, by assigning a unique chemical species to each image. Figure \ref{fig:concept_storage} depicts this multiple input storage format for $M$ binary image inputs.

To construct data in this parallel format, we need to obtain a set of compatible chemical compounds. In general, the following criteria must be met by all species in a chemical set to be considered valid for the proposed data storage scheme:

\begin{enumerate}
    \item The chemicals must be miscible in the chosen solvent.
    \item The chemicals should be stable, relatively inert, and should not react with one another.
    \item The chemicals should be compatible with analytical chemistry tools that can quantify their concentrations.
\end{enumerate}

\begin{table}[b]
\renewcommand{\arraystretch}{2}
\centering
\begin{center}
\caption{Computational cost of classifying $M$ binary inputs, each containing $N$ bits, in a traditional versus volumetric perceptron}
\begin{tabular}{|c|c|c|}
\hline                       
\textbf{Operations} & \textbf{\begin{tabular}{@{}c@{}}Scalar Single \\[-1em] Core Silicon\end{tabular}} & \textbf{\begin{tabular}{@{}c@{}}Parallel Chemical \\[-1em] Mixtures\end{tabular}} \\\hline
Additions & $M{\cdot}N-1$ & \multirow{2}{*}{$N$}  \\ \cline{1-2}
Multiplications & $M{\cdot}N$ & \\ \hline     
Total & $2{\cdot}M{\cdot}N - 1$ & $N$ \\ \hline 
\end{tabular}\label{tab:operations}
\end{center}
\end{table}

The potential advantages of computing with chemical mixtures stem from the ability for many datasets to coexist in parallel. For instance, in the case of overlaid chemical images, any operation on a single well will simultaneously be applied to the corresponding pixel in all images. As such, this encoding scheme has the potential to support massively parallel storage and computation. Table \ref{tab:operations} shows a comparison of the number of operations required for a perceptron with a traditional computer versus the proposed mixture-based technique. The number of operations needed to be performed with chemical mixtures scales only with the number of input features and is independent of the number of input instances.

\subsection{Computing with Chemical Mixtures} \label{ssec:computation}

\begin{figure}[t]
\centerline{\includegraphics[scale=0.22]{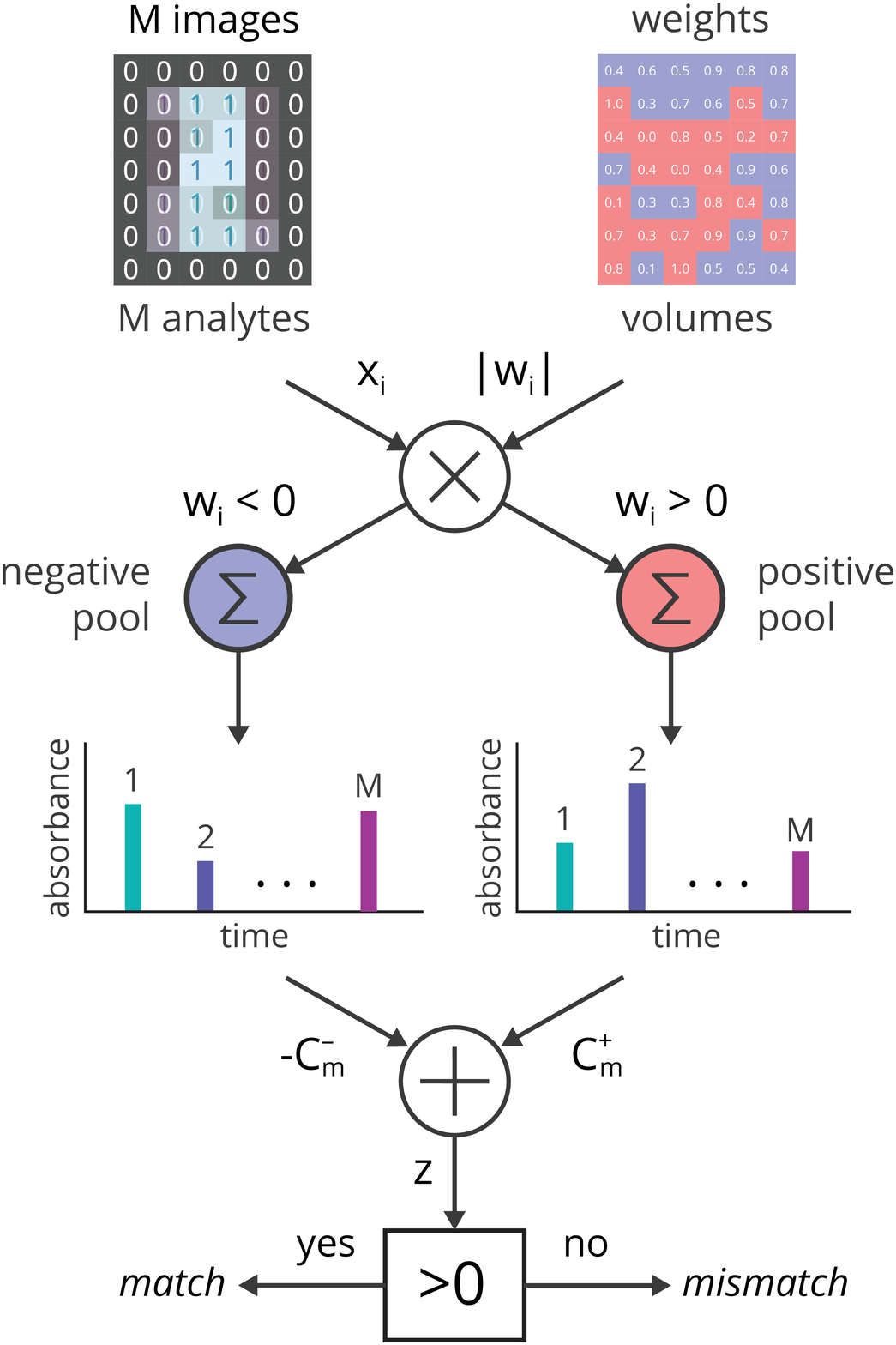}}
\caption{A schematic of the proposed chemical computation procedure, as implemented for pattern classification. All spatially concurrent chemical datasets ($x$) are operated on in parallel by a single weight matrix ($w$), whose values are realized as volumetric fluid transfers. Since weights can be positive and negative ($w_i\in[-1,1]$), a pool for each polarity is made. Each pool is analyzed by liquid chromatography to measure the concentrations of each analyte species. The differential concentration of each analyte is calculated in post-processing and used to determine the appropriate label for the input data.}
\label{fig:concept_compute}
\end{figure}

Figure \ref{fig:concept_compute} illustrates the computational scheme for the proposed chemical mixture based perceptron. The perceptron weights ($w_i\in [-1,1]$) are scaled to correspond to a maximum volume $V_o$, which is chosen based on the available volume in the data wells. Since we can only transfer positive liquid volumes, we pool wells with positive and negative weights in two separate MAC operations. 

The total volume that will be transferred from the $i^{th}$ well will be: $V_i = |w_i| \cdot V_o$. As previously described, the scaling of the transfer volume represents a multiplication and the pooling of volumes into a common well represents an addition. Since bits from different datasets may be stored in the same well, these pooling operations allow for parallel multiply-accumulate operations on all concurrently stored datasets. There is zero marginal computational cost to increasing parallelism, since, regardless of the complexity of the chemical mixtures, we only need to perform the pooling transfers once.

To show that the system shown in Figure \ref{fig:concept_compute} realizes the perceptron classifier, it is instructive to work backwards from the output of the system. We can write the output for the data represented by molecular species $m$ as:

\begin{equation}
	z_m = \Delta C_m = C_m^+ - C_m^-
	\label{eq:explicitoutput}
\end{equation}

\noindent where $C_m^+$ and $C_m^-$ are the concentrations of species $m$ in the positive and negative weight pools, respectively. According to Equation \ref{eq:mixconcentration} the concentration of molecule $m$ at the output of each MAC can be expressed as:

\begin{equation}
C_m^+ = \sum_{\substack{i=1\\w_i>0}}^N \frac{V_i}{V_p^+} \cdot C_{mi} = \sum_{\substack{i=1\\w_i>0}}^N \frac{|w_i| \cdot V_o}{V_p^+} \cdot C_{mi}
\end{equation}

\noindent and similarly:

\begin{equation}
C_m^- = \sum_{\substack{i=1\\w_i<0}}^N \frac{|w_i| \cdot V_o}{V_p^-} \cdot C_{mi}
\end{equation}

\noindent where $V_p^+$ and $V_P^-$ are the total volumes in each pool, $i$ is the index of the data well, $V_i = |w_i|\cdot V_o$ is the weighted volume transferred from the $i^{th}$ well, and $C_{mi}$ is the concentration of molecule $m$ in the $i^{th}$ well. We can then expand Equation \ref{eq:explicitoutput} as:

\begin{equation}
	z_m = \sum_{\substack{i=1\\w_i>0}}^N \frac{|w_i|\cdot V_o}{V_p^+} \cdot C_{mi} - \sum_{\substack{i=1\\w_i<0}}^N \frac{|w_i|\cdot V_o}{V_p^-} \cdot C_{mi}
\end{equation}

\noindent As long as the the pooled volumes are intentionally set to be equal after weighted pooling ($V_p^+=V_p^-=V_p$), by appropriately adding pure solvent, we can collect the summations as:

\begin{equation}
	z_m = \sum_{i=1}^N \frac{w_i\cdot V_o}{V_p} \cdot C_{mi} = \sum_{i=1}^N w_i \cdot x_{mi}
\end{equation}

\noindent where our features have been defined to be the scaled data concentrations: $x_{mi} = \frac{V_o}{V_p} \cdot C_{mi}$. This yields the original form of the pre-classification output that we sought to generate.

\subsection{Reading the Results of Chemical Mixture Computations}

\begin{figure}[t!]
\centerline{\includegraphics[scale=0.28]{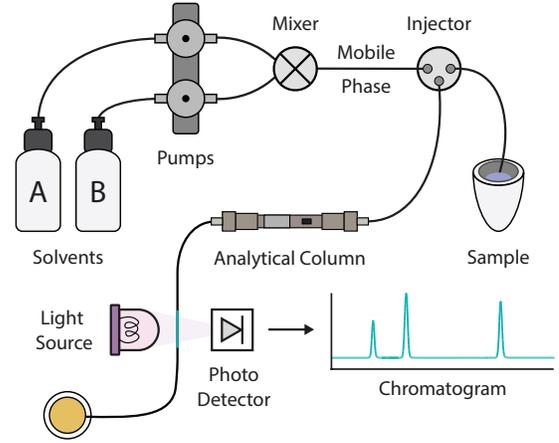}}
\caption{High Performance Liquid Chromatography (HPLC). A liquid sample is injected into a stream of solvents. Together, the sample and solvents are forced, at high pressure, through an analytical column. Depending on the type of column and solvents, some chemicals will exit the column at different times. At the end of the column is an ultraviolet (UV) light source and a photodetector. If an appropriate wavelength is selected, the analyte can be detected by a change in absorbance as it exits the column. The absorbance is plotted over time as a chromatogram.}
\label{fig:HPLC}
\end{figure}

To verify the output of the computations, we need to determine the amount of each component present in the liquid samples. For this purpose, we chose to employ high performance liquid chromatography (HPLC). HPLC is a technique commonly used in analytical chemistry for separation, identification, and quantification of components in a mixture \cite{karger1997hplc}. An HPLC is comprised of an injection port, several pumps, an ultraviolet light source, and a photodetector, as illustrated in Figure \ref{fig:HPLC}. A liquid sample is injected into a stream of solvents known as the mobile phase. The mobile phase, now containing the sample to be analyzed, flows through a column containing a solid adsorbent known as the stationary phase. Components in the sample interact with the adsorbent to varying degrees based on their chemical properties, causing different chemicals to leave the column (to `elute'), at different times. These retention times can be used to identify individual components. When a compound that absorbs UV light elutes from the column and passes the detector, a peak in optical absorbance is observed, and the area of this peak is proportional to the relative abundance of the compound in the sample. The resulting absorbance time series, known as a chromatogram, can be used to both identify and quantify the chemicals in a mixture.

\section{System Development}
\label{sec:development}

\begin{figure}[t]
\centerline{\includegraphics[scale=0.19]{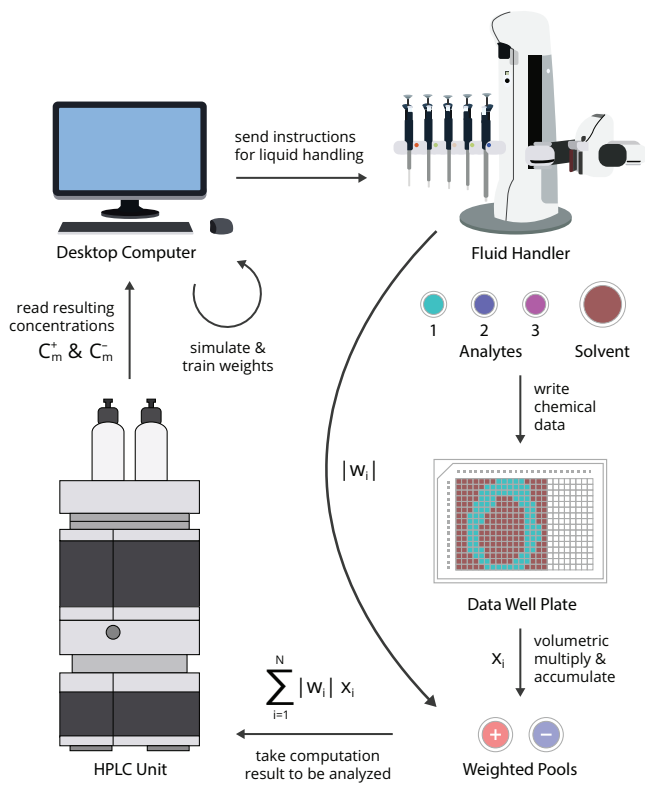}}
\caption{An overview of the experimental setup and data flow used for these experiments. Weight matrices were trained in simulation and then converted, along with test data, into sequences of pipetting instructions for a robotic liquid handler. Analytes were dispensed into a 384-well microplate to form the chemical dataset and then collected in volume fractions corresponding to the classifier weight matrix. The outputs were analyzed by HPLC to produce class labels.}
\label{fig:setup}
\end{figure}

\subsection{Experimental Setup}

A diagram of our experimental setup and procedural workflow is shown in Figure \ref{fig:setup}. Binary input data and classifier weights are first trained through simulations on a desktop computer, in a Python environment. Prior to chemically encoding the input data, concentrated stock solutions are prepared of each analyte (selected analytes are described in Subsection \ref{sec:analyte}), at 62.5\,mg/mL in dimethyl sulfoxide (DMSO, Macron Fine Chemicals 4948-02). To write the chemical data to a 384-well plate, the binary datasets are converted to pipetting instructions for a fluid handling robot (Andrew Alliance 1000G). For each input data bit whose value is `1', the robot is instructed to transfer $20{\,\mu}L$ of the corresponding analyte to the appropriate well. If the input data is `0', it transfers $20{\,\mu}L$ of solvent (DMSO) instead. After the chemical datasets are created, the classifier weights are converted into additional pipetting instructions which the robot uses to perform the weighted-summation operations, placing the pooled outputs into an empty well plate. Once the positive and negative weight pools are generated, each output is analyzed using high performance liquid chromatography. The analytes representing each dataset exit the instrument at different times, allowing separate estimations of the output concentration of each component. For each analyte, the differential concentration (${\Delta}C_m$) is calculated on a computer. If it is greater than zero, then the data contained in the well plate is classified as a match; otherwise, the data is classified as mismatch.

\subsection{Chemical Selection}
\label{sec:analyte}

We selected three similar phenol compounds to encode the data in our experiments: 2,4,6-tri-tert-butylphenol (analyte 1, 98\%, Sigma Aldrich T49409), 2,6-dimethylphenol (analyte 2, $\geq$99.5\%, Sigma Aldrich D174904), and 4-nitrophenol (analyte 3, $\geq$99\%, Sigma Aldrich 241326). Phenols were selected due to their established UV absorbance \cite{mateos2001, montedoro1992} and well-defined HPLC peak shapes. We initially also considered benzoates and other aromatics. However, phenol compounds also offer a range of chemical functionalization options through acylation \cite{murashige2011comparisons}, protection with benzyl \cite{kuwano2008benzyl} or tert-butyloxycarbonyl \cite{cheraiet2013simple}, silylation \cite{sefkow1999selective}, or even methylation \cite{ouk2002methylation}. This introduces the possibility of increasing parallelism through simple chemical manipulations, which would generate numerous related compounds with similar UV sensitivity.

\begin{figure}[t]
\centerline{\includegraphics[scale=0.29]{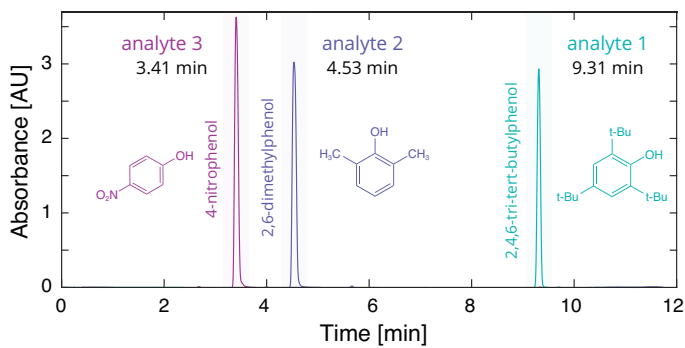}}
\caption{HPLC chromatograms for the three phenols used to encode data in this study. The chemical concentrations in these three samples were 7\,mg/mL for analyte 1, 5\,mg/mL for analyte 2, and 8.5\,mg/mL for analyte 3. The absorbances were measured at 214\,nm. The elution time can be used to identify the analyte, while the area under each peak can be related to the amount of the compound in the sample.}
\label{fig:chroma_time}
\end{figure}

\subsection{Measurement Calibration}

The HPLC used in this work was an Agilent 1260 infinity series model with a quaternary pump, a standard autosampler, a thermostatted column compartment, and a variable wavelength detector that was set at 204, 214, and 254\,nm with bandwidths of 4\,nm. To identify the characteristic elution time of each analyte, the three compounds were measured independently using a C18 reversed phase column (Agilent Poroshell 120 EC-C18, 699975-902, 4.5\,mm x 50\,mm, 2.7$\,\mu$m particle size). A gradient flow program was run with a two-part mobile phase comprised of water (A) and acetonitrile (B), where the volume ratios at each gradient time endpoint were set to: 95\%-A \& 5\%-B at 0\,min, 5\%-A \& 95\%-B at 9\,min, 5\%-A \& 95\%-B at 10\,min, 95\%-A \& 5\%-B at 11\,min, and 95\%-A \& 5\%-B at 12.5\,min. Chromatograms for the three analytes are shown in Figure \ref{fig:chroma_time}. The elution times were determined to be 3.41 minutes, 4.53 minutes, and 9.31 minutes, and the pulse width for each species was approximately 9 seconds. We note that while our proof of concept experiments use only 3 analytes, the narrow pulse width implies that as many as 80 analytes could be independently quantified using this HPLC protocol.

\begin{figure}[t]
\centerline{\includegraphics[scale=0.27]{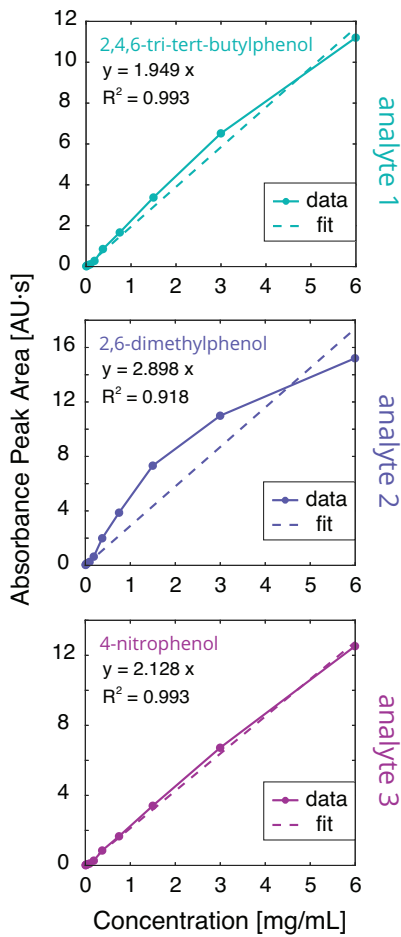}}
\caption{HPLC concentration calibration curves for the three analytes used in this study. Using the known elution time of each compound, we measured and herein plot the area under the absorbance peak for each analyte at various concentrations.} 
\label{fig:chroma_conc}
\end{figure}

In Figure \ref{fig:chroma_conc}, we present calibration curves for the three analytes, which relate the true concentration to the area under the measured HPLC chromatogram peak. A concentrated equimolar mixture (12\,mg/mL of each of the three analytes) was prepared and serially diluted to obtain samples with varying concentrations. Using 11 serial 2:1 dilutions, the concentrations were varied from 6\,mg/mL to 0.006\,mg/mL. Each of the diluted samples was analyzed with HPLC, and the areas of the chromatogram peaks were calculated. For each compound, we performed a simple zero-intercept linear fit to relate peak area and compound abundance. We note that although photodetector saturation causes noticeable nonlinearity at higher concentrations, our technique relies on differential concentrations, and thus achieving the correct classification output only requires that the detector output be monotonic.

\section{Experiments \& Results}
\label{sec:experiments}

\subsection{MNIST Image Classification}

\begin{figure}[t]
\centerline{\includegraphics[scale=0.28]{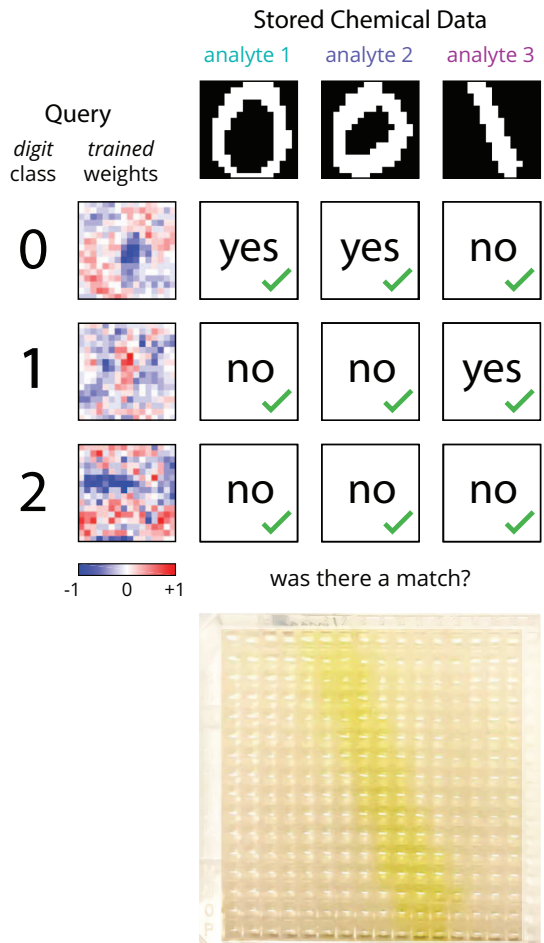}}
\caption{Chemical classification of MNIST handwritten digits. Three 16$\times$16 (256-bit) binary images were chemically encoded, in parallel, on a 384-well plate. The overlaid chemical images were then classified by three different perceptrons which had been previously trained to identify either digit `0', `1', or `2'. The results of this experiment are shown in a table format as class matches ($z_m>0$) or mismatches ($z_m<0$). All nine chemical classifier outputs were correct (3 true positives, 6 true negatives). A photograph of the microplate containing the chemical dataset of overlaid images is also shown. Each well in the plate contains $60{\,\mu}$L of liquid whose chemical composition represents the values of one pixel across three images.}
\label{fig:largerimage}
\end{figure}

For a first experimental demonstration, we used images derived from the well-known MNIST database of handwritten digits \cite{mnist}. The original images were grayscale at 28$\times$28 pixel resolution, but for these experiments, we binarized and resized the images to 16$\times$16. We trained three one-versus-all classifiers on a computer \emph{a priori} for three foreground classes, representing the digits `0', `1', and `2'. Each classifier was trained using 100 foreground class images and 100 background class images which were randomly selected from the MNIST training set. For example, the classifier with the digit `0' foreground class was trained using 100 images of the digit `0' and 100 images of other digits `1' through `9'. Color maps of the trained weight matrices are shown in Figure \ref{fig:largerimage}. 

We constructed a dataset of three overlaid MNIST images, consisting of two distinct `0' images and one image of `1'. These images were mapped onto a well plate and encoded with the three previously discussed analytes. The resulting microplate is shown in Figure \ref{fig:largerimage}, where the chemically encoded images are faintly visible due to the colors of the analyte solutions (particularly analyte 3). We used the three perceptron classifiers to operate on this chemical data, and the resulting MNIST classifications are shown in Figure \ref{fig:largerimage}. As expected, the `0' classifier correctly identified the two images with zeros, and the `1' classifier correctly identified the image of a one. In total, all 9 of the MNIST perceptron outputs were correctly labeled. We note that while these perceptrons performed well, the exact accuracy of the classifiers is not the main focus of this paper. Rather, our aim is to to reproduce the perceptron operations using chemical computations.

\subsection{Performance Evaluation}

\begin{figure}[t]
\centerline{\includegraphics[scale=0.52]{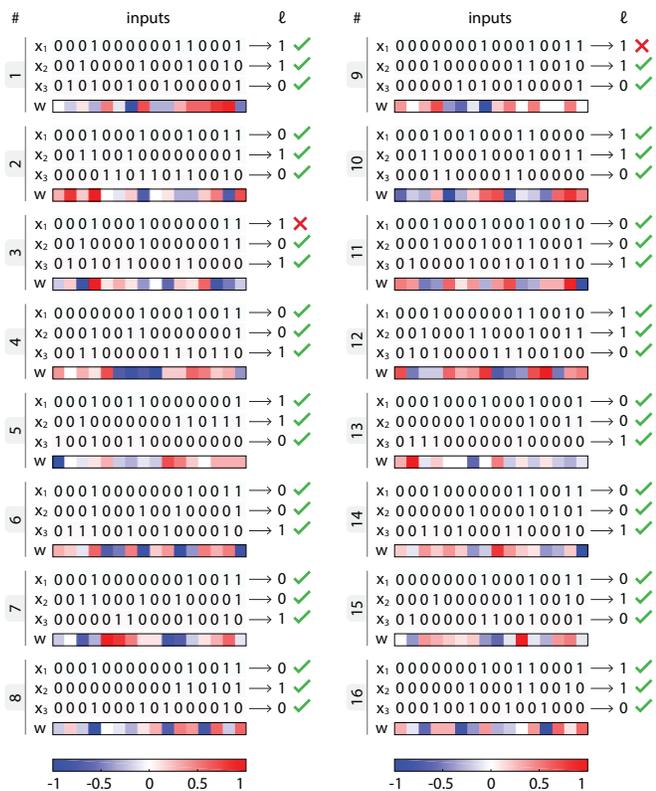}}
\caption{Validation experiments for chemical classifiers with pseudo-random data. Sixteen trials were performed. In each trial, three 16-bit data vectors ($x_1,x_2,x_3$) were chemically encoded and classified according to a weight vector ($w$). The computed class label ($\ell$) is shown for each vector, along with a green check mark or red cross out to indicate whether or not the chemical classifier identified it correctly. In total, 46 out of 48 vectors were correctly classified (96\% accurate with 2 false positives).}
\label{fig:validationdata}
\end{figure}

Our chemical computation is not limited to images and is extensible to linear classification of any binary dataset. To evaluate the robustness of the computations, we performed a set of experiments using smaller pseudo-random binary vectors. Sixteen 16-element weight vectors ($w \in [-1,1]$) were selected at random, as shown in Figure \ref{fig:validationdata}. For each $w$, we chose three 16-bit data vectors, selected such that one vector is classified with large margin as a mismatch ($\ell=0$), one vector is easily classified as a match ($\ell=1$), and one vector is near the classifier's boundary.

\begin{figure}[t]
\centerline{\includegraphics[scale=0.25]{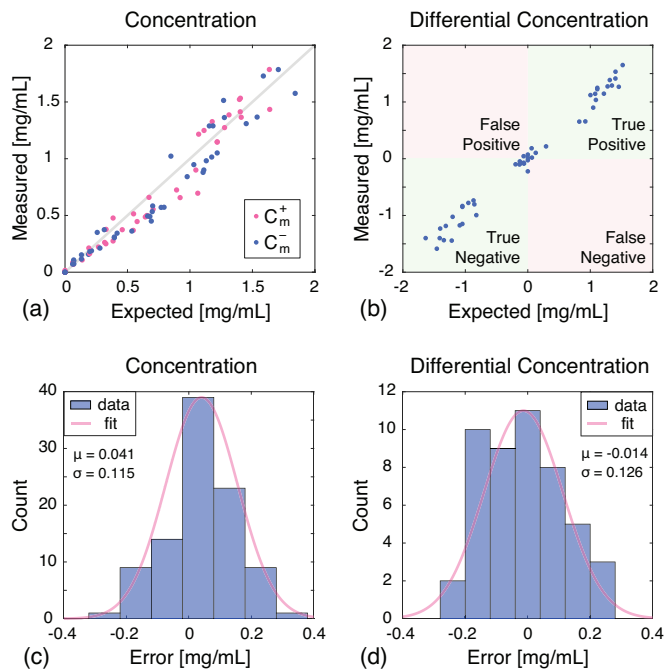}}
\caption{Performance assessment of the chemical multiply-accumulate operations carried out in the sixteen validation trials (48 data vectors). (a) Expected and measured analyte concentrations in the positive and negative chemical output pools. (b) A scatter plot of expected and measured differential concentrations ($C_m^+ - C_m^-$). In total, 46 out of the 48 vectors were correctly classified. The two misclassified vectors occurred for small differential concentrations and were both false positives. (c) A histogram of the differences between expected and measured absolute concentration. (d) A histogram of the differences between expected and measured differential output concentrations. Both histograms include a fit to a normal distribution.}
\label{fig:performance}
\end{figure}

The expected and HPLC-measured concentrations of the positively and negatively weighted pools are shown in Figure \ref{fig:performance}. The expected and measured values of the differential concentration are also shown. In both cases, deviations from a straight line ($y=x$) represent errors in the chemical encoding, computation, or measurement. In Figure \ref{fig:performance}b, points which appear in the upper left quadrant are false positives, while points which appear in the lower right quadrant are false negatives. Histograms of these errors are also shown.

The classification output is robust to moderate experimental variations, but the exact output concentrations are affected by several sources of experimental variability. Inconsistent pipetting volumes during the initial dataset creation, as well as the tolerance of the weighted summing volumes, will cause variations in the pooled output. Additionally, the HPLC is sensitive to small changes in its sample injection volume, and the chromatogram calibration can drift slightly over time. In total, we observed errors on the scale of 10\% of the expected outputs, which naturally has a larger affect on decisions closer to the classifier boundary. The mean differential concentration error was close to zero (0.041\,mg/mL), and the 3$\,\sigma$ spread was approximately 0.3\,mg/mL. The overall chemical classifier accuracy was 96\%, correctly classifying 46 out of 48 test vectors.

\section{Discussion and Future Directions}
\label{sec:future}

The promise of this approach as a valuable alternative computing model hinges on its ability to scale up to operating in parallel on many datasets. Currently, the scale of our demonstrations is limited by the throughput of the automated liquid handler. For example, between assembling the dataset and performing the classifications, the MNIST demonstration required a total of 1,716 liquid transfers, used 697 disposable pipette tips, and took 24 hours to complete. While this is a considerable amount of time, it does not represent a physical limit, as higher throughput liquid handling systems are common in pharmaceutical laboratories. Moving forward, we anticipate that improved robotics will allow us to increase the computational throughput by several orders of magnitude.

The parallelism of the chemical analysis is another important avenue for improvement. Our demonstrations used a set of only three compounds, but the current system could be readily scaled to several dozen. This would not impact the inference or readout time, but the time to prepare the input data wells would increase. Putting aside the dataset creation time, scaling up to hundreds or thousands of parallel computations would likely require supplementing HPLC with other analytical techniques such as mass spectrometry \cite{zeng1998, shockcor1996}.

The finite volume of the chemical datasets implies a limit to the number of times that data can be sampled and processed. In the MNIST experiments, each well began with a volume of 60$\,\mu$L, and the classifier weights were scaled to transfer between 0\,$-$\,6.25$\,\mu$L from each well. Thus the current system can operate on each data well approximately 20 times. With improved liquid handling tools, the sample volumes can be reduced to make more efficient use of each chemical dataset.

Our current demonstrations require analyzing two chemical output pools and electronically calculating the differential concentrations of their analytes. If we are to scale up the scope and complexity of our chemical computational system, we will need to develop approaches for cascading multiple computing stages and integrating other operations beyond linear volumetric additions and multiplications. More sophisticated reaction networks have been developed for DNA-based neural networks \cite{DNA-winner-take-all-NN}, but the broader space of non-biological chemical reaction networks remains sparsely explored, and may yield rich new opportunities for chemical computing platforms. 


\section{Conclusion} 
\label{sec:conclusion}

We have presented a scheme for implementing linear classification operations using chemical mixtures. Binary input data is encoded in the chemical composition of an array of liquid samples, and a robotic fluid handler is programmed to perform multiplications and additions as fractional volume transfers and pooling operations. The chemical coding enables parallel computation and allows for increased information density. The result of the volumetric operations is represented in the concentration of chemicals in output pools, which are analyzed using high performance liquid chromatography. We used this system for parallel classification of several 16$\times$16 binary MNIST images of handwritten digits, as well as a set of pseudo-random binary vectors. The method's overall accuracy was demonstrated, producing 55 correct classifications out of 57 tests. 

While these demonstrations are limited, we consider this work to be a first step towards a new class of chemical-domain computational engines which can operate on increasingly parallel datasets. We anticipate that chemical computation will find applications in ultra-low-power systems, extreme environments, and as complements to electronic computing systems \cite{adleman}. In a fashion similar to how DNA archival data storage may soon complement traditional electronic media \cite{dna-next-gen-storage, dna-practical-storage, dna-archival-storage}, and neural-inspired computing has revolutionized the way we process large datasets \cite{NNimagenet, NNspeech, NNface, Tann2018}, chemical-domain computation may pave the road to entirely new ways of leveraging the information processing capabilities of the natural world.


\section*{Acknowledgement}

The authors would like to thank Joseph Geiser, Peter Weber, Jason Sello, Rukshan Perera, and Eamonn Kennedy for many helpful discussions.

\bibliographystyle{IEEEtran}
\bibliography{references}

\begin{thebibliography}{10}
\providecommand{\url}[1]{#1}
\csname url@samestyle\endcsname
\providecommand{\newblock}{\relax}
\providecommand{\bibinfo}[2]{#2}
\providecommand{\BIBentrySTDinterwordspacing}{\spaceskip=0pt\relax}
\providecommand{\BIBentryALTinterwordstretchfactor}{4}
\providecommand{\BIBentryALTinterwordspacing}{\spaceskip=\fontdimen2\font plus
\BIBentryALTinterwordstretchfactor\fontdimen3\font minus
  \fontdimen4\font\relax}
\providecommand{\BIBforeignlanguage}[2]{{%
\expandafter\ifx\csname l@#1\endcsname\relax
\typeout{** WARNING: IEEEtran.bst: No hyphenation pattern has been}%
\typeout{** loaded for the language `#1'. Using the pattern for}%
\typeout{** the default language instead.}%
\else
\language=\csname l@#1\endcsname
\fi
#2}}
\providecommand{\BIBdecl}{\relax}
\BIBdecl

\bibitem{alberts2017}
B.~Alberts, \emph{Molecular Biology of the Cell}.\hskip 1em plus 0.5em minus
  0.4em\relax Garland Science, 2017.

\bibitem{deepNNoverview}
J.~Schmidhuber, ``Deep learning in neural networks: An overview,'' \emph{Neural
  Networks}, vol.~61, pp. 85--117, 2015.

\bibitem{goelet1986}
P.~Goelet, V.~F. Castellucci, S.~Schacher, and E.~R. Kandel, ``The long and the
  short of long--term memory----a molecular framework,'' \emph{Nature}, vol.
  322, no. 6078, p. 419, 1986.

\bibitem{RoseISIT}
C.~Rose, S.~Reda, B.~Rubenstein, and J.~K. Rosenstein, ``Computing with
  chemicals: Perceptrons using mixtures of small molecules,'' in \emph{2018
  IEEE International Symposium on Information Theory (ISIT)}, Jun. 2018, pp.
  2236--2240.

\bibitem{Kennedy2018}
E.~Kennedy, P.~Shakya, M.~Ozmen, C.~Rose, and J.~K. Rosenstein,
  ``Spatiotemporal information preservation in turbulent vapor plumes,''
  \emph{Applied Physics Letters}, vol. 112, no.~26, p. 264103, 2018.

\bibitem{molecularneuron}
W.~T. Huang, L.~X. Chen, J.~L. Lei, H.~Q. Luo, and N.~B. Li, ``Molecular
  neuron: {From} sensing to logic computation, information encoding, and
  encryption,'' \emph{Sensors and Actuators B: Chemical}, vol. 239, pp.
  704--710, 2017.

\bibitem{molecularsensing}
S.~A. Salehi, H.~Jiang, M.~D. Riedel, and K.~K. Parhi, ``Molecular {Sensing}
  and {Computing} {Systems},'' \emph{IEEE Transactions on Molecular, Biological
  and Multi-Scale Communications}, vol.~1, no.~3, pp. 249--264, 2015.

\bibitem{dna-fountain}
Y.~Erlich and D.~Zielinski, ``{D}{N}{A} fountain enables a robust and efficient
  storage architecture,'' \emph{Science}, vol. 355, no. 6328, pp. 950--954,
  2017.

\bibitem{dna-crispr-cas-encoding}
S.~L. Shipman, J.~Nivala, J.~D. Macklis, and G.~M. Church,
  ``{C}{R}{I}{S}{P}{R}--{C}as encoding of a digital movie into the genomes of a
  population of living bacteria,'' \emph{Nature}, vol. 547, no. 7663, p. 345,
  2017.

\bibitem{dna-rand-access}
L.~Organick, S.~D. Ang, Y.-J. Chen, R.~Lopez, S.~Yekhanin, K.~Makarychev, M.~Z.
  Racz, G.~Kamath, P.~Gopalan, B.~Nguyen \emph{et~al.}, ``Random access in
  large-scale {D}{N}{A} data storage,'' \emph{Nature Biotechnology}, vol.~36,
  no.~3, p. 242, 2018.

\bibitem{perceptron}
F.~Rosenblatt, ``The perceptron: A probabilistic model for information storage
  and organization in the brain.'' \emph{Psychological Review}, vol.~65, no.~6,
  p. 386, 1958.

\bibitem{perceptron1}
M.~Minsky, S.~A. Papert, and L.~Bottou, \emph{Perceptrons: An Introduction to
  Computational Geometry}.\hskip 1em plus 0.5em minus 0.4em\relax MIT Press,
  2017.

\bibitem{prml}
C.~Bishop, \emph{Pattern Recognition and Machine Learning}.\hskip 1em plus
  0.5em minus 0.4em\relax Springer, 2006.

\bibitem{multilayerperceptron}
G.~J. Gibson and C.~F. Cowan, ``On the decision regions of multilayer
  perceptrons,'' \emph{Proceedings of the IEEE}, vol.~78, no.~10, pp.
  1590--1594, 1990.

\bibitem{SVMnonlinear}
Z.~Fu, A.~Robles-Kelly, and J.~Zhou, ``Mixing linear {S}{V}{M}s for nonlinear
  classification,'' \emph{IEEE Transactions on Neural Networks}, vol.~21,
  no.~12, pp. 1963--1975, 2010.

\bibitem{taga2003}
M.~E. Taga and B.~L. Bassler, ``Chemical communication among bacteria,''
  \emph{Proceedings of the National Academy of Sciences}, vol. 100, no. suppl
  2, pp. 14\,549--14\,554, 2003.

\bibitem{bloom1984}
F.~Bloom, ``The functional significance of neurotransmitter diversity,''
  \emph{American Journal of Physiology-Cell Physiology}, vol. 246, no.~3, pp.
  C184--C194, 1984.

\bibitem{dna-next-gen-storage}
G.~M. Church, Y.~Gao, and S.~Kosuri, ``Next-generation digital information
  storage in {D}{N}{A},'' \emph{Science}, p. 1226355, 2012.

\bibitem{dna-archival-storage}
J.~Bornholt, R.~Lopez, D.~M. Carmean, L.~Ceze, G.~Seelig, and K.~Strauss, ``A
  {D}{N}{A}-based archival storage system,'' \emph{ACM SIGOPS Operating Systems
  Review}, vol.~50, no.~2, pp. 637--649, 2016.

\bibitem{karger1997hplc}
B.~L. Karger, ``{H}{P}{L}{C}: Early and recent perspectives,'' \emph{Journal of
  Chemical Education}, vol.~74, no.~1, p.~45, 1997.

\bibitem{mateos2001}
R.~Mateos, J.~L. Espartero, M.~Trujillo, J.~Rios, M.~Le{\'o}n-Camacho,
  F.~Alcudia, and A.~Cert, ``Determination of phenols, flavones, and lignans in
  virgin olive oils by solid-phase extraction and high-performance liquid
  chromatography with diode array ultraviolet detection,'' \emph{Journal of
  Agricultural and Food Chemistry}, vol.~49, no.~5, pp. 2185--2192, 2001.

\bibitem{montedoro1992}
G.~Montedoro, M.~Servili, M.~Baldioli, and E.~Miniati, ``Simple and
  hydrolyzable phenolic compounds in virgin olive oil. 1. {T}heir extraction,
  separation, and quantitative and semiquantitative evaluation by
  {H}{P}{L}{C},'' \emph{Journal of Agricultural and Food Chemistry}, vol.~40,
  no.~9, pp. 1571--1576, 1992.

\bibitem{murashige2011comparisons}
R.~Murashige, Y.~Hayashi, S.~Ohmori, A.~Torii, Y.~Aizu, Y.~Muto, Y.~Murai,
  Y.~Oda, and M.~Hashimoto, ``Comparisons of {O}-acylation and
  {F}riedel--{C}rafts acylation of phenols and acyl chlorides and {F}ries
  rearrangement of phenyl esters in trifluoromethanesulfonic acid: Effective
  synthesis of optically active homotyrosines,'' \emph{Tetrahedron}, vol.~67,
  no.~3, pp. 641--649, 2011.

\bibitem{kuwano2008benzyl}
R.~Kuwano and H.~Kusano, ``Benzyl protection of phenols under neutral
  conditions: Palladium-catalyzed benzylations of phenols,'' \emph{Organic
  Letters}, vol.~10, no.~10, pp. 1979--1982, 2008.

\bibitem{cheraiet2013simple}
Z.~Cheraiet, S.~Hessainia, S.~Ouarna, M.~Berredjem, and N.-E. Aouf, ``A simple
  and eco-sustainable method for the {O}-{B}oc protection/deprotection of
  various phenolic structures under water-mediated/catalyst-free conditions,''
  \emph{Green Chemistry Letters and Reviews}, vol.~6, no.~3, pp. 211--216,
  2013.

\bibitem{sefkow1999selective}
M.~Sefkow and H.~Kaatz, ``Selective protection of either the phenol or the
  hydroxy group in hydroxyalkyl phenols,'' \emph{Tetrahedron Letters}, vol.~40,
  no.~36, pp. 6561--6562, 1999.

\bibitem{ouk2002methylation}
S.~Ouk, S.~Thiebaud, E.~Borredon, P.~Legars, and L.~Lecomte, ``{O}-methylation
  of phenolic compounds with dimethyl carbonate under solid/liquid phase
  transfer system,'' \emph{Tetrahedron Letters}, vol.~43, no.~14, pp.
  2661--2663, 2002.

\bibitem{mnist}
Y.~LeCun, L.~Bottou, Y.~Bengio, and P.~Haffner, ``Gradient-based learning
  applied to document recognition,'' \emph{Proceedings of the IEEE}, vol.~86,
  no.~11, pp. 2278--2324, 1998.

\bibitem{zeng1998}
L.~Zeng and D.~B. Kassel, ``Developments of a fully automated parallel
  {H}{P}{L}{C}/mass spectrometry system for the analytical characterization and
  preparative purification of combinatorial libraries,'' \emph{Analytical
  Chemistry}, vol.~70, no.~20, pp. 4380--4388, 1998.

\bibitem{shockcor1996}
J.~P. Shockcor, S.~E. Unger, I.~D. Wilson, P.~J. Foxall, J.~K. Nicholson, and
  J.~C. Lindon, ``Combined {H}{P}{L}{C}, {N}{M}{R} spectroscopy, and ion-trap
  mass spectrometry with application to the detection and characterization of
  xenobiotic and endogenous metabolites in human urine,'' \emph{Analytical
  Chemistry}, vol.~68, no.~24, pp. 4431--4435, 1996.

\bibitem{DNA-winner-take-all-NN}
K.~M. Cherry and L.~Qian, ``Scaling up molecular pattern recognition with
  {D}{N}{A}-based winner-take-all neural networks,'' \emph{Nature}, vol. 559,
  no. 7714, p. 370, 2018.

\bibitem{adleman}
L.~M. Adleman, ``Molecular computation of solutions to combinatorial
  problems,'' \emph{Science}, vol. 266, no. 5187, pp. 1021--1024, 1994.

\bibitem{dna-practical-storage}
N.~Goldman, P.~Bertone, S.~Chen, C.~Dessimoz, E.~M. LeProust, B.~Sipos, and
  E.~Birney, ``Towards practical, high-capacity, low-maintenance information
  storage in synthesized {D}{N}{A},'' \emph{Nature}, vol. 494, no. 7435, p.~77,
  2013.

\bibitem{NNimagenet}
A.~Krizhevsky, I.~Sutskever, and G.~E. Hinton, ``Imagenet classification with
  deep convolutional neural networks,'' in \emph{Advances in Neural Information
  Processing systems}, 2012, pp. 1097--1105.

\bibitem{NNspeech}
A.~Graves, A.~Mohamed, and G.~Hinton, ``Speech recognition with deep recurrent
  neural networks,'' in \emph{Acoustics, Speech and Signal Processing (ICASSP),
  2013 IEEE International Conference on}.\hskip 1em plus 0.5em minus
  0.4em\relax IEEE, 2013, pp. 6645--6649.

\bibitem{NNface}
S.~Lawrence, C.~L. Giles, A.~C. Tsoi, and A.~D. Back, ``Face recognition: A
  convolutional neural-network approach,'' \emph{IEEE Transactions on Neural
  Networks}, vol.~8, no.~1, pp. 98--113, 1997.

\bibitem{Tann2018}
\BIBentryALTinterwordspacing
H.~Tann, S.~Hashemi, and S.~Reda, ``Flexible deep neural network processing,''
  \emph{arXiv}, 2018. [Online]. Available:
  \url{http://arxiv.org/abs/1801.07353}
\BIBentrySTDinterwordspacing

\end{thebibliography}
\end{document}